# The effective hydrodynamic radius is not a constant

Gan Ren (任淦)[†]

Departments of Physics & Key laboratory of photonic and optical detection in civil aviation, Civil Aviation Flight University of China, Guanghan 618307, China

**Abstract**

The effective hydrodynamic radius is usually assumed to be a constant in testing the Stokes-Einstein relation by its variants. We have performed molecular dynamics simulations with ortho-terphenyl and Kob-Andersen model to examine the assumption and found the effective hydrodynamic radius is not a constant but decreases with decreasing temperature. The variant of Stokes-Einstein relation is not established but Stokes-Einstein relation is valid by considering the changes of the effective hydrodynamic radius. We propose the assumption should be seriously evaluated when using variants to test Stokes-Einstein relation.

## 1. Introduction

The Stokes-Einstein (SE) relation[1] $D = k_B T / C\eta a$ combines the Einstein relation $D = k_B T / \alpha$ and Stokes' formula $\alpha = C\eta a$, where $D$ is the diffusion coefficient, $k_B$ is the Boltzmann constant, $T$ is the temperature, $\alpha$ is the frictional coefficient, $\eta$ is the viscosity, $a$ is the effective hydrodynamic radius and $C$ is a constant depending on the boundary condition. SE relation has been successfully applied to many situations, such as the protein diffusion[2] and the oxygen transport.[3] However, it is proposed to be invalid for liquids undergo supercooling.[4-6] Instead of the original formula $D = k_B T / C \eta a$, the validity of SE relation in supercooled liquids is usually tested by the three variants, $D \sim T/\eta$,[4, 7-10] $D \sim \tau^{-1}$,[6, 11-16] and $D \sim T/\tau$,[5, 17-19], where $\tau$ is the structural relaxation time. The variant $D \sim T/\eta$ is based on the assumption that $a$ is a constant under different conditions. Both the variant $D \sim \tau^{-1}$ and $D \sim T/\tau$ are based on $D \sim T/\eta$. The $D \sim \tau^{-1}$ is based on a further assumption that $\tau$ has a similar temperature dependence as $\eta/T$. The $D \sim T/\tau$ is based on another approximate relation $\eta = G_\infty \tau$, where $G_\infty$ is the instantaneous shear modulus presumed to be a constant.

*This work was supported by the Science Foundation of Civil Aviation Flight University of China (Nos. J2021-054, JG2019-19).
[†]E-mail address: rengan@alumni.itp.ac.cn

The three variants are all based on the assumption that $a$ is a constant under different conditions. Its rationality is important on the conclusion drawn on the testing of SE relation. Actually, there exists some evidences suggest that $a$ is varied with conditions. A classical case is the deviations from SE relation for alkali ions in aqueous solutions.[20, 21] The $a$ deviates from the crystallographic radius especially for the smallest Li$^+$ due to the dielectric polarization, which is even larger than that of Cs$^+$. It is almost proportional to the volume fraction in a diluted organic aqueous solution.[22] The gyration radius is often adopted to estimate $a$ in aqueous macromolecule solutions, but the gyration radius is dependent on conditions.[23]

Another example suggests $a$ changes with conditions is existed such system that the molecule (or ion) with smaller mass and size has a smaller $D$ than that with larger mass and size at a certain temperature (labeled as $T_1$), such as NaCl aqueous solution,[24, 25] [EMI$^+$][NO$_3^-$][26] and [EMI$^+$][BF$_4^-$][27] ionic liquids at room temperature. Considering such a system with constant volume under heating and according to the gas kinetic theory,[28] the smaller molecule will even have a larger $D$ than the larger at high enough temperature (labeled as $T_2$). Because the system is in equilibrium, the SE relation is valid for $T_1$ and $T_2$; we get $a_{l1} < a_{s1}$ at $T_1$ and $a_{l2} > a_{s2}$ at $T_2$; where subscript '1' and '2' corresponding to temperature $T_1$ and $T_2$ separately, subscripts '$l$' and '$s$' label the larger and the smaller molecule separately. The result indicates $a$ has to be varied with temperature.

The mechanism behind is that molecule is not free but interacts with its surrounding molecules. Molecule moves partially with its surrounding shells and behaves as a composite particle. The interaction will increase relatively for supercooled liquids under cooling for the increasing ratio of the potential energy to thermodynamic kinetic energy. In addition, the SE relation as a special case of the fluctuation-dissipation theorem is established in equilibrium and near equilibrium state. The supercooled liquids are still in equilibrium and the SE relation should be valid.[29] Based on the above facts, we have enough reason to believe $a$ is varied with conditions especially for supercooled liquids. As the importance of $a$ for the SE relation, it is necessary to examine the rationality of the assumption. In this work, molecular dynamics (MD) simulations were performed with two classical supercooled liquid models to test the rationality of the assumption.

## 2. Simulation details and analysis methods

Our MD simulations are performed with the two classical supercooled liquid models, ortho-terphenyl (OTP) and Kob-Andersen model, to verify above proposition. The simulation details are same as previous studies, the OTP system[5] contains 3072 molecules with a constant density 1.0746g/cm$^3$ and the Kob-Andersen system[30] contains 8000 particles (A:B = 80:20) in a cubic box with size 6.392nm. All simulations were performed in *NVT* ensemble with the GROMACS package.[31, 32] The simulated temperature range is within 260-400K with 13 temperatures for OTP and is within 66-500K (0.66-5.0 in reduced unit) with 24 temperatures for Kob-Andersen. The three variants of SE relation for both models are all observed to be invalid in the chosen temperature range.[5] The system temperature was kept a constant by the Nosé-Hoover thermostat.[33, 34] The van der Waals interactions were calculated directly with the cutoff of 2.0 and 0.85 nm for OTP and Kob-Andersen, respectively.

The method proposed by Hess[35] is adopted to calculate $\eta$ for its reliability and fast convergence. An external force $a_x = A \cdot \cos(qz)$ with maximum $A$ is applied in the $X$ direction, where $q = 2\pi/l$ with $l$ the box size. The shear viscosity is described by $\eta = A\rho/Vq^2$ with $V$ the maximum of the velocity in the $X$ direction and $\rho$ the density. Because of the same $\rho$ and $q$ for all simulations, we use ratio *A/V* to evaluate the shear viscosity and which is in unit ps$^{-1}$. The $\alpha$ is determined by introducing a small force $f_e$ to a part of particles in the linear response regime. The particle B is mainly considered for Kob-Andersen model in this work; 160 OTP molecules and 400 B particles are separately chosen for OTP and Kob-Andersen to keep enough statistical accuracy and avoid too much disturbance on the system. After getting the steady state, we have $f_e = \alpha v$ for particle with velocity $v$. Because of the same mass *m* for each dragged particle, the ratio $f_e m/v$ is adopted to evaluate the frictional coefficient $\alpha$ and which is in unit ps$^{-1}$. The diffusion coefficient is calculated via its asymptotic relation with the mean square displacement $D = \lim_{t \to \infty} \langle \Delta \mathbf{r}^2(t) \rangle / 6t$, where $\Delta \mathbf{r}(t)$ is particle position displacement and $\langle \rangle$ denotes ensemble average.

## 3. Results and discussion

The viscosity $\eta$, frictional coefficient $\alpha$ and diffusion constant *D* at different temperature *T* for OTP and Kob-Andersen are calculated and plotted in Fig. 1. To examine the assumption, we have rescaled the $\eta$ and $\alpha$ by the value at *T* = 400 K for the OTP and at *T* = 500 K for Kob-Andersen, respectively. If the *a* is a constant, the $\bar{\eta}$ and $\bar{\alpha}$ will be fallen on the line $\bar{\alpha} = \bar{\eta}$ shown in Fig. 2a. However, it deviates from the line $\bar{\alpha} = \bar{\eta}$ and the deviations become larger as $\bar{\eta}$ and $\bar{\alpha}$ increase with decreasing temperature. Moreover, the deviations are inclined more to the $\bar{\eta}$ axis as with a larger $\bar{\eta}$. The results indicate *a* is not a constant but decreases as temperature decreases.

Moreover, $\alpha$ and $\eta$ usually follow Arrhenius law as $\alpha = \alpha_0 e^{E_\alpha/T}$ and $\eta = \eta_0 e^{E_\eta/T}$, respectively.[36] However, there is no reason to believe that the two will have the same activation energies, namely $E_\alpha = E_\eta$, to guarantee *a* not changing with conditions. We further evaluate the assumption by analyzing the scaling relation between $\alpha$ and $\eta$ by $\alpha \sim \eta^\beta$. Fig. 2b shows the exponent $\beta$ is 0.88 for OTP and 0.83 for Kob-Andersen, both deviate from 1.0 over 0.1. It also indicates the $a$ is not a constant but decreases as temperature decreases. Combined the results given by Figs. 2a and 2b, both the deviations from $\bar{\alpha}=\bar{\eta}$ and $\beta=1.0$ indicate that the assumption of the $a$ is a constant is not valid for both OTP and Kob-Andersen, and which becomes smaller with decreasing temperature. Similar results as shown by $a \sim T/D\eta$ or $D \sim (\eta/T)^{-\xi}$ with $\xi<1$ are also observed in other supercooled liquids, including supercooled water,[11] supercooled binary Lennard-Jones liquids,[10, 37] water/methanol solutions,[9] and tris-Naphthylbenzene.[38]

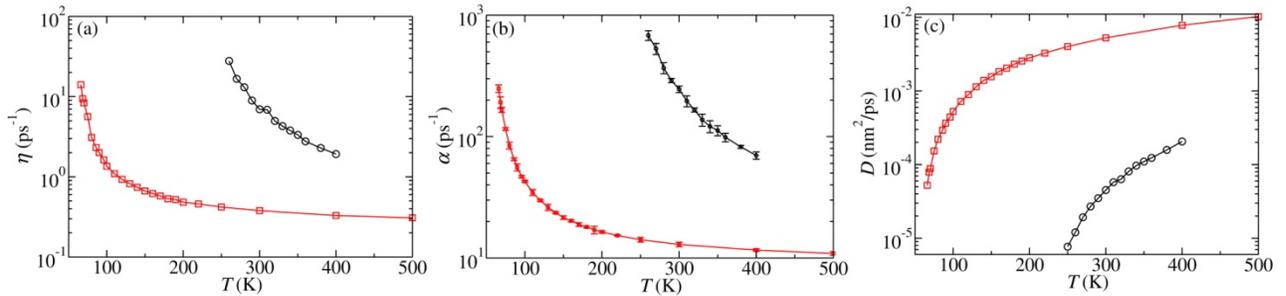

**Fig. 1.** The $\eta$ vs $T$ (a), $\alpha$ vs $T$ (b) and $D$ vs $T$ (c) for OTP and Kob-Andersen. The data for OTP is colored in black and the red is for Kob-Andersen.

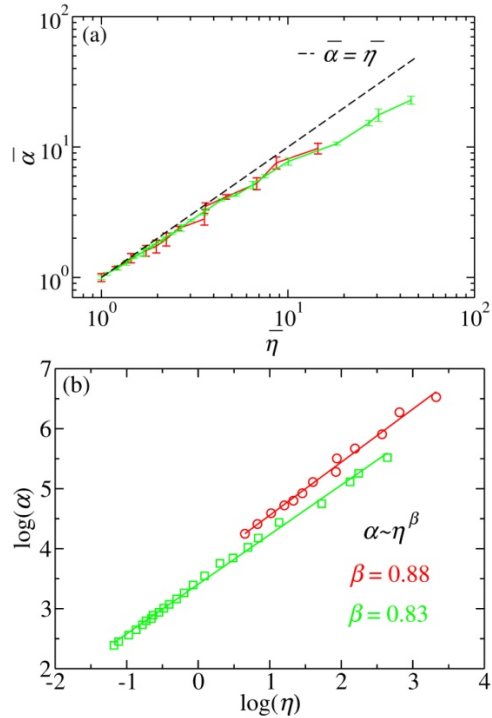

**Fig. 2.** The rescaled viscosity $\bar{\eta}$ vs rescaled frictional coefficient $\bar{\alpha}$ (a) and the scaling relation

$\alpha \sim \eta^\beta$ (b) for OTP and Kob-Andersen model. The data for OTP are in red and Kob-Andersen in green. The symbols in (b) are simulated results and the solid are fitting lines.

Because of the importance of the assumption, it is necessary to take the changes of $a$ into account when testing SE relation. Because the two variants, $D \sim \tau^{-1}$ and $D \sim T/\tau$, are based on $D \sim T/\eta$, we mainly compare the original SE relation $D = k_B T/\alpha$ with the variant $D \sim T/\eta$ to see the differences considering the changes of $a$ or not. The logarithm of $D$ and $T/\eta$, $T/\alpha$ for OTP and Kob-Andersen are plotted in Fig. 3. The variant $D \sim T/\eta$ behave as a fractional form $D \sim (\eta/T)^{-\xi}$ with $\xi \simeq 0.9$ for both OTP and Kob-Andersen as shown by Figs. 3a and 3c. The results are similar as previous studies observed in other liquids such as supercooled water,[7] aqueous NaCl solutions[25] and ionic liquids.[39] Taking $a$ into account, Figs. 3b and 3d show the original SE relation $D \sim T/\alpha$ is valid for both OTP and Kob-Andersen that the exponent $\xi$ in $D \sim (\alpha/T)^{-\xi}$ is very close to 1.0. Combined the results given by $D \sim T/\eta$ and $D \sim T/\alpha$, it is shown that SE relation given by variant $D \sim T/\eta$ is not established but SE relation $D \sim T/\alpha$ is valid after considering the changes of $a$. The results indicate the $a$ is important for the conclusions draw on the validity of SE relation and should be seriously considered to avoid giving some ambiguous conclusions.

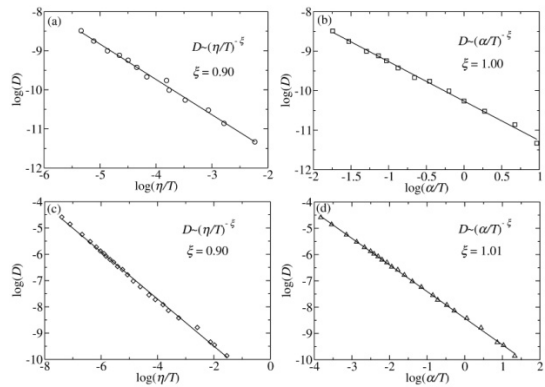

**Fig. 3.** Verification of the validities of the SE relation $D \sim T/\alpha$ and its variant $D \sim T/\eta$ for OTP and Kob-Andersen model: (a) $D \sim T/\eta$ and (b) $D \sim T/\alpha$ for OTP; (c) $D \sim T/\eta$ and (d) $D \sim T/\alpha$ for Kob-Andersen. The calculated data are represented by symbols and the solid lines are fitted by

$$D \sim (\eta/T)^{-\xi} \text{ and } D \sim (\alpha/T)^{-\xi},$$ respectively.

## 4. Conclusions

In summary, we have examined the assumption of the effective hydrodynamic radius $a$ is a constant in testing the SE relation by performing MD simulations with OTP and Kob-Andersen model. Our results indicate the assumption is invalid, and $a$ is not a constant but decreases with decreasing temperature. The SE relation given by variant $D \sim T/\eta$ is breakdown and in fractional form, however, it is actually established after taking the changes of $a$ into account. It is shown that the $a$ is an important parameter to test the validity of SE relation and some ambiguous conclusions may be drawn when using the assumption arbitrarily. So we propose the assumption should be seriously evaluated when using the variants to test the SE relation.

## Acknowledgments

The computations of this work were conducted on the Tian-2 supercomputer.